\documentstyle[twoside,fleqn,espcrc2,epsf]{article}

\newcommand{\be}{\begin{equation}}
\newcommand{\ee}{\end{equation}}
\newcommand{\bdm}{\begin{displaymath}}
\newcommand{\edm}{\end{displaymath}}
\newcommand{\<}{\langle}
\renewcommand{\>}{\rangle}

\def\su3{$SU(3)$}

\def\spose#1{\hbox to 0pt{#1\hss}}
\def\ltapprox{\mathrel{\spose{\lower 3pt\hbox{$\mathchar"218$}}
 \raise 2.0pt\hbox{$\mathchar"13C$}}}
\def\gtapprox{\mathrel{\spose{\lower 3pt\hbox{$\mathchar"218$}}
 \raise 2.0pt\hbox{$\mathchar"13E$}}}
\def\inapprox{\mathrel{\spose{\lower 3pt\hbox{$\mathchar"218$}}
 \raise 2.0pt\hbox{$\mathchar"232$}}}

\title{
Topology and chiral symmetry in finite temperature
QCD\thanks{Presented by U.~M.~Heller at {\sl Lattice '99.}}
}

\author{
Robert G.~Edwards\address{
SCRI, The Florida State University, 
Tallahassee, FL 32306-4130, USA},
Urs M.~Heller$^{\rm a}$,
Joe Kiskis\address{
Dept. of Physics, University of California,
Davis, CA 95616, USA}
and Rajamani Narayanan\address{
American Physical Society,
One Research Road,
Ridge, NY 11961, USA}
}

\begin{document}

\begin{abstract}
We investigate the realization of chiral symmetry in the vicinity of
the deconfinement transition in quenched QCD using overlap fermions.
Via the index theorem obeyed by the overlap fermions, we gain insight
into the behavior of topology at finite temperature.
We find small eigenvalues, clearly separated from the bulk of the
eigenvalues, and study the properties of their distribution.
We compare the distribution with a model of a dilute gas of instantons
and anti-instantons and find good agreement.

\end{abstract}

\maketitle


\section{ Some properties of the overlap Dirac operator}

The overlap Dirac operator is constructed out of a Wilson-like Dirac
operator (the usual Wilson Dirac operator is used for this talk)
with a {\sl large negative mass} \cite{Neuberg98,ehn_pbp}:
\be
D(\mu)=\frac{1}{2} \left[1+\mu + (1-\mu) \gamma_5\epsilon(H_w(m))  \right] .
\ee
Here $H_w(m) = \gamma_5 D_{\rm Wilson}(-m)$. This form insures that
topology is seen by the overlap fermions, since $Q={\rm tr}\epsilon(H_w)/2$.
The quark mass is proportional to $\mu$ ($-1 < \mu <1$) for small $\mu$.

The propagator for external fermions is given by
\be
{\tilde D}^{-1}(\mu) = (1-\mu)^{-1} \left[ D^{-1}(\mu) -1 \right] ,
\ee
{\it i.e.} it has a contact term subtracted. The propagator is then
chiral in the massless case.

In many cases, it is more convenient to use the hermitian version
$H_o(\mu) = \gamma_5 D(\mu)$. The massless version satisfies,
\be
\{H_o(0), \gamma_5 \} = 2 H_o^2(0) .
\ee
This is the Ginsparg-Wilson relation, which is often used these days to prove
good chiral properties of the overlap Dirac operator.

It follows that $[H_o^2(0), \gamma_5] = 0$, {\it i.e.} the eigenvectors
of $H_o^2(0)$ can be chosen as chiral. Since
\be
H_o^2(\mu) = ( 1 - \mu^2 ) H_o^2(0) + \mu^2 ,
\ee
this holds also for the massive case.

\vskip -12cm
\rightline{FSU-SCRI-99C-49}
\vskip +11.6cm

Each eigenvalue $0 < \lambda^2 < 1$ of $H_o^2(0)$ is doubly degenerate
with opposite chirality eigenvectors. In this basis, $H_o(\mu)$ and $D(\mu)$
are $2 \times 2$ block diagonal. In $D(\mu)$ the blocks are
\be
\pmatrix{
         (1-\mu) \lambda^2 + \mu & (1-\mu) \lambda \sqrt{1-\lambda^2} \cr
         -(1-\mu) \lambda \sqrt{1-\lambda^2} & (1-\mu) \lambda^2 + \mu
  }
\ee
where
\be
\gamma_5 = \pmatrix{ 1 & 0 \cr 0 & -1 } .
\ee

For a gauge field with topological charge $Q \ne 0$, there are, in addition,
$|Q|$ exact zero modes with chirality ${\rm sign}(Q)$ paired with
eigenvectors of opposite chirality and eigenvalue 1. These are also
eigenvectors of $H_o(\mu)$ and $D(\mu)$:
\be
D(\mu)_{\rm zero~sector} : \quad \pmatrix{
                         \mu & 0 \cr 0 & 1 }
  \quad {\rm or} \quad
  \pmatrix{ 1 & 0 \cr 0 & \mu }
\ee
depending on the sign of $Q$.

\section{ Small eigenvalues and the chiral condensate}

In the chiral eigenbasis of $H_o^2(0)$, the external propagator takes the
block diagonal form with $2 \times 2$ blocks
\begin{eqnarray}
{\tilde D}^{-1}(\mu) : && (\lambda^2 (1-\mu^2) + \mu^2)^{-1} \times
 \\ \nonumber
  &&     \pmatrix{
         \mu (1-\lambda^2) & -\lambda \sqrt{1-\lambda^2} \cr
         \lambda \sqrt{1-\lambda^2} & \mu (1-\lambda^2)
  } .
\end{eqnarray}
In topologically non-trivial background fields, there are $|Q|$ additional
blocks
\be
\pmatrix{ \frac{1}{\mu} & 0 \cr 0 & 0 }
  \qquad {\rm or} \qquad
\pmatrix{ 0 & 0 \cr 0 & \frac{1}{\mu} }
\ee
depending on the sign of $Q$.

We thus find in a fixed gauge field background,
\be
\< \bar \psi \psi \>(\{U\}) = \frac{|Q|}{\mu V} 
 + \frac{1}{V} \sum_{\lambda > 0} \frac{2\mu (1-\lambda^2)}
 {\lambda^2 (1-\mu^2) + \mu^2} ,
\ee
and averaged over gauge fields, we get the condensate.
It is dominated by the small (non-zero) eigenvalues, and in the thermodynamic
limit where the first term vanishes, it is given by the density of
eigenvalues at zero $\rho(0^+)$.

The close connection between the distribution of small eigenvalues and
the condensate motivated us to study the small eigenvalues of the overlap
Dirac operator in quenched QCD as the temperature is raised above the
deconfining transition temperature $T_c$. A similar study for $T=0$ can
be found in \cite{Ov_RMT}.

\section{ Small eigenvalue distribution in quenched QCD above $T_c$}

\begin{figure}
\vskip 3mm
\epsfxsize=2.5in
\centerline{\epsfbox[105 90 571 546]{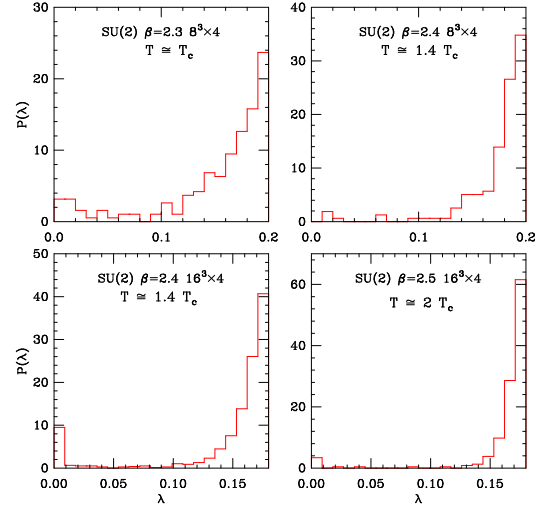}}
\vskip -4mm
\caption{Small eigenvalue distribution for various SU(2) ensembles for
temperatures above $T_c$ on lattices with $N_t=4$.}
\label{fig:su2_dist}
\end{figure}

\begin{figure}
\vskip -4mm
\epsfysize=4.0in
\centerline{\epsfbox[75 90 601 616]{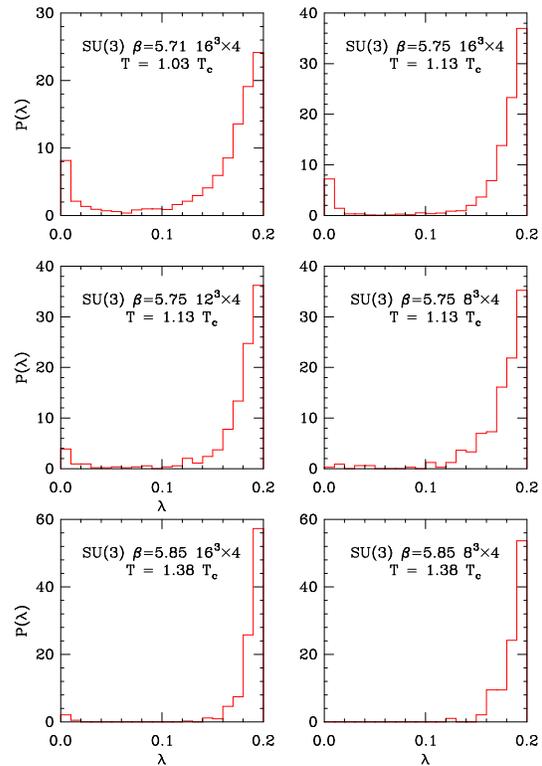}}
\caption{Same as Fig.~\protect\ref{fig:su2_dist} but for SU(3).}
\label{fig:su3_dist}
\end{figure}

\begin{figure}
\epsfxsize=2.5in
\centerline{\epsfbox[85 50 591 556]{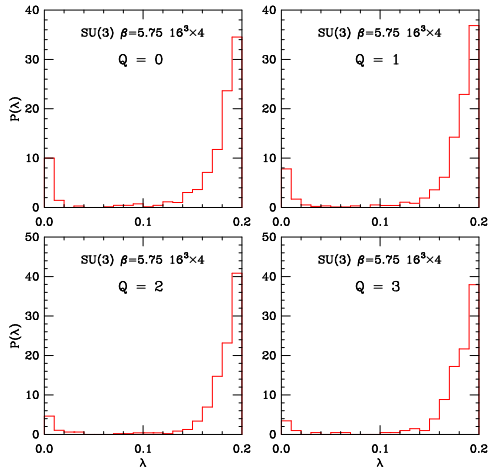}}
\vskip -8mm
\caption{Small eigenvalue distribution in various topological charge
sectors for SU(3) with $\beta=5.75$ on a $16^3\times 4$ lattice.}
\vskip -5mm
\label{fig:su3_dist_Q}
\end{figure}

We focus on the non-zero eigenvalues and display their distributions
for SU(2) and SU(3) at various temperatures and spatial volumes
in Figs.~\ref{fig:su2_dist}--\ref{fig:su3_dist_Q}. All figures show
a region around $\lambda=0.05$ with very few eigenvalues, or even
none. There is also a concentration of small eigenvalues below this value
except in the cases with the smallest volumes and the largest $\beta$'s.
Fig.~\ref{fig:su3_dist_Q} shows that these very small non-zero

In the remainder, we shall focus on the small modes $\lambda < 0.05$.
We denote the number of zero and small nonzero eigenvalues with chirality
$\pm$ by $n_\pm$ and the total number by $n=n_++n_-$. The topological
charge is given by $Q=n_+-n_-$. Our results for the various ensembles
are given in Tables~\ref{tab:su2_ev} and \ref{tab:su3_ev}.
We see that for fixed $\beta$, $\<n\>/V$ and $\<Q^2\>/V$ seem to remain
finite and nonzero in the large volume limit, but they drop quickly as
$\beta$, and hence the temperature, is increased.

\section{ Modeling by a dilute instanton -- anti-instanton gas}

Looking in more detailed at the small modes, we find that their number
$n$ is roughly Poisson distributed $ P(n,\<n\>) = \<n\>^n e^{-\<n\>}/n!$,
and in particular, the average and variance are approximately equal. Also
for fixed $n$, $n_+$ and $n_-$ are roughly binomially distributed.
These observations are consistent with interpreting the small modes to
be due to a dilute gas of instantons and anti-instantons with $n_+$ and
$n_-$ their numbers. $n-|Q|$ of the would-be zero modes overlap and mix
to give small eigenvalues, while $|Q|$ exact zero modes remain.

\begin{table}
\begin{center}
\vskip -5mm
\caption{
$\<n\>$ and $\<Q^2\>$ per $8^3$ spatial volume and  $\<n\>/\sigma_n$ with
$\sigma_n$ the variance of $n$, for our SU(2) ensembles.}
\label{tab:su2_ev}
\vspace*{0.1cm}
\begin{tabular}{c|cc|cc} \hline
 volume           &\multicolumn{2}{|c|}{$8^3\times 4$}
                  &\multicolumn{2}{|c}{$16^3\times 4$} \\
 $\beta$          & 2.3  & 2.4  & 2.4  & 2.5   \\ \hline
 $\<n\>/V$        & 1.7  & 0.29 & 0.25 & 0.054 \\
 $\<Q^2\>/V$      & 1.7  & 0.31 & 0.25 & 0.053 \\
 $\<n\>/\sigma_n$ & 0.98 & 1.09 & 0.93 & 0.99 \\ \hline
\end{tabular}
\vskip -8mm
\end{center}
\end{table}

\begin{table}
\begin{center}
\vskip -5mm
\caption{Same as Table \protect\ref{tab:su2_ev} but for SU(3).}
\label{tab:su3_ev}
\vspace*{0.1cm}
\begin{tabular}{c|cc|c|ccc} \hline
 volume           &\multicolumn{2}{|c|}{$8^3\times 4$} & $12^3\times 4$
                  &\multicolumn{3}{|c}{$16^3\times 4$} \\
 $\beta$          & 5.75 & 5.85  & 5.75 & 5.71 & 5.75 & 5.85  \\ \hline
 $\<n\>/V$        & 0.32 & 0.058 & 0.28 & 0.63 & 0.31 & 0.051 \\
 $\<Q^2\>/V$      & 0.31 & 0.068 & 0.28 & 0.64 & 0.33 & 0.049 \\
 $\<n\>/\sigma_n$ & 1.05 & 0.90  & 0.92 & 1.15 & 1.02 & 0.83  \\ \hline
\end{tabular}
\vskip -8mm
\end{center}
\end{table}

At finite temperature, instantons fall off exponentially, and so do the
fermionic zero modes associated with them. We consider a toy model of
randomly (Poisson and binomially) distributed instantons and
anti-instantons, inducing interactions of the form
$h_0 {\rm e}^{-d(i,j)/D}$ between the would-be zero modes of every instanton
-- anti-instanton pair $(i,j)$ with separation $d(i,j)$. Like sign pairs
are assumed to have no interactions. This toy model reproduces the
qualitative features of the small eigenvalue distributions. Preliminary
estimates give $D\approx 1/(2T)$~\cite{Ov_FT}.

In conclusion, we find that in quenched QCD above the deconfining transition
temperature topology, as manifested by exact zero modes, persists. Furthermore,
a finite density of small eigenvalues, separated from the bulk of the
eigenvalues, remains. The properties of their distribution is well modeled
by a dilute gas of instantons and anti-instantons.
Our observation might be a quenched artefact, since in full QCD the finite
temperature transition is believed to be driven by chiral symmetry
restoration.


This research was supported by DOE contracts 
DE-FG05-85ER250000 and DE-FG05-96ER40979.
Computations were performed on the QCDSP and CM-2 at SCRI,
and on workstations at UC Davis and SCRI.

\vskip -3mm

\end{document}